\documentclass[aps,prd,preprintnumbers,amsmath,amssymb,twocolumn,floatfix,superscriptaddress,nofootinbib]{revtex4-1}

\usepackage{amssymb,amsmath}
\usepackage{mathtools}
\usepackage{bm}
\usepackage{graphicx}
\usepackage{epstopdf}
\usepackage{hyperref}
\usepackage{array}
\usepackage{soul}
\usepackage{braket}
\usepackage{float}
\usepackage[usenames, dvipsnames]{color}
\usepackage{slashed}
\usepackage{csquotes}

\usepackage{soul,xcolor}
\setstcolor{blue}

\enlargethispage{\baselineskip}

\hypersetup{
     colorlinks   = true,
     citecolor    = blue
}

\begin{document}

\newcommand*{\bfrac}[2]{\genfrac{}{}{0pt}{}{#1}{#2}}
\newcommand{\tos}{t_{\rm osc}}
\newcommand{\Ho}{H_{\rm osc}}
\newcommand{\ho}{h_{\rm osc}}
\newcommand{\co}{\chi_{\rm osc}}
\newcommand{\mP}{m_{\rm Pl}}
\newcommand{\THi}{\tau_{\rm Higgs}}
\newcommand{\THu}{\tau_{\rm Hubble}}
\newcommand{\TRH}{T_{\rm reh}}
\newcommand{\hh}{\langle h^2\rangle}
\newcommand{\LL}{\mathcal{L}}
\newcommand{\WW}{\mathcal{W}_k}
\newcommand{\HH}{\mathcal{H}}
\newcommand{\hrms}{h_{\rm rms}}
\newcommand{\td}{t_{\rm reh}}
\newcommand{\tdo}{t_{\rm reh,0}}
\newcommand{\Nr}{N_{\rm reh}}

\newcommand{\es}[1]{ \textcolor{cyan}{[{\bf ES}: #1]}}
\newcommand{\ps}[1]{ \textcolor{blue}{[{\bf PS}: #1]}}
\newcommand{\lv}[1]{\textcolor{RawSienna}{[{\bf LV}: #1]}}
\newcommand{\edit}[1]{\textcolor{orange}{ #1}}
\newcommand{\esedit}[1]{\textcolor{cyan}{ #1}}

\title{
Primordial non-Gaussianity from the Effects of the Standard Model Higgs during Reheating after Inflation
}

\newcommand{\FIRSTAFF}{\affiliation{The Oskar Klein Centre for Cosmoparticle Physics,
	Department of Physics,
	Stockholm University,
	AlbaNova,
	10691 Stockholm,
	Sweden}}
\newcommand{\SECONDAFF}{\affiliation{Department of Physics,
	University of Texas, Austin,
	Texas 78712, USA}}
\newcommand{\THIRDAFF}{\affiliation{Nordita,
	KTH Royal Institute of Technology and Stockholm University
	Roslagstullsbacken 23,
	10691 Stockholm,
	Sweden}}
\newcommand{\FOURTHAFF}{\affiliation{Institut de F{\'i}sica d'Altes Energies (IFAE), The Barcelona Institute of Science and Technology (BIST), Campus UAB, 08193 Bellaterra, Barcelona}}
\newcommand{\FIFTHAFF}{\affiliation{Nikhef,
	Science Park 105,
	1098 XG Amsterdam, The Netherlands}}	
\newcommand{\SIXTHAFF}{\affiliation{Institute Lorentz of Theoretical Physics,
	University of Leiden, 2333CA Leiden, The Netherlands}}
\newcommand{\SISSA}{\affiliation{Scuola Internazionale Superiore di Studi Avanzati (SISSA), via Bonomea 265, 34136 Trieste, Italy}}
\newcommand{\INFNTrieste}{\affiliation{INFN, Sezione di Trieste, via Valerio 2, 34127 Trieste, Italy}}
\newcommand{\IFPU}{\affiliation{Institute for Fundamental Physics of the Universe (IFPU), via Beirut 2, 34151 Trieste, Italy}}
\newcommand{\TDLI}{\affiliation{Tsung-Dao Lee Institute (TDLI),
    520 Shengrong Road, 201210 Shanghai, P.\ R.\ China}}
\newcommand{\SJTU}{\affiliation{School of Physics and Astronomy, Shanghai Jiao Tong University,
    800 Dongchuan Road, 200240 Shanghai, P.\ R.\ China}}

\author{Aliki Litsa}
\email[Electronic address: ]{aliki.litsa@fysik.su.se}
\FIRSTAFF

\author{Katherine Freese}
\email[Electronic address: ]{ktfreese@utexas.edu}
\FIRSTAFF
\SECONDAFF
\THIRDAFF

\author{Evangelos I. Sfakianakis}
\email[Electronic address: ]{esfakianakis@ifae.es}
\FOURTHAFF
\FIFTHAFF
\SIXTHAFF

\author{Patrick Stengel}
\email[Electronic address: ]{pstengel@fe.infn.it}
\FIRSTAFF

\SISSA
\INFNTrieste
\IFPU

\author{Luca Visinelli}
\email[Electronic address: ]{luca.visinelli@sjtu.edu.cn}
\TDLI\SJTU

\preprint{NORDITA-2020-083; UTTG-17-2020; Nikhef-2020-037}
\date{\today}

\begin{abstract}
We propose a new way of studying the Higgs potential at extremely high energies. The Standard Model (SM) Higgs boson, as a light spectator field during inflation in the early Universe, can acquire large field values from its quantum fluctuations which vary among different causal (Hubble) patches. Such a space dependence of the Higgs after the end of inflation leads to space-dependent SM particle masses and hence variable efficiency of reheating, when the inflaton decays to Higgsed SM particles. Inhomogeneous reheating results in (observable) temperature anisotropies. Further, the resulting temperature anisotropy spectrum acquires a significant non-Gaussian component, which is constrained by \textit{Planck} observations of the Cosmic Microwave Background (CMB) and potentially detectable in next-generation experiments. Constraints on this non-Gaussian signal largely exclude the possibility of the observed temperature anisotropies arising primarily from Higgs effects. Hence, in principle, observational searches for non-Gaussianity in the CMB can be used to constrain the dynamics of the Higgs boson at very high (inflationary) energies.
\end{abstract}
\maketitle

\section{Introduction}
\label{sec:Introduction}

Inflation, an early period of accelerated expansion, was proposed to explain the homogeneity, isotropy, and flatness of the Universe~\cite{Guth:1980zm, Sato:1980yn, Brout:1977ix}. A simple inflationary mechanism consists of a scalar field, the inflaton, rolling down a nearly flat potential that dominates the energy density of the Universe~\cite{Linde:1981mu, Albrecht:1982wi}. Quantum fluctuations of the inflaton field give rise to density perturbations that can seed the formation of large scale structures in the Universe including galaxies and clusters. Temperature anisotropies in the Cosmic Microwave Background (CMB) sourced by these density fluctuations provide among the strongest observational probes of inflation. 

After inflation, the Universe must transition into the radiation dominated era via a reheating mechanism, during which the inflaton decays into light Standard Model (SM) particles or an intermediate sector. These decays can either occur perturbatively~\cite{Dolgov:1982th, Abbott:1982hn} or lead to resonant particle production~\cite{Greene:1997fu, Chung:1999ve}. If the reheating process is inhomogeneous, it provides a second mechanism for generating density perturbations (in addition to those described above); these can also seed the growth of structure and produce observable anisotropies in the CMB~\cite{Dvali:2003ar, Dvali:2003em, Kofman:2003nx,Kobayashi:2011hp,DeSimone:2012qr,Langlois:2013dh,Karam:2020skk}.

The standard inflationary paradigm, in which a single inflaton field slowly rolls down a flat potential, results in perturbations with a highly Gaussian probability distribution. CMB observations to date are consistent with Gaussianity; indeed, the \textit{Planck} satellite has placed significant bounds on non-Gaussianity (NG) that already rule out many non-standard models of inflation. A future detection of NG could challenge this paradigm and teach us about the nature of inflation. Inflationary scenarios resulting in significant NG include those with multiple fields~\cite{Bernardeau:2002jy}, non-Bunch-Davies initial conditions~\cite{Holman:2007na, Kundu:2011sg}, non-canonical kinetic terms~\cite{Garcia-Saenz:2019njm}, or non-linear growth of perturbations after inflation~\cite{Scoccimarro:1995if, Scoccimarro:1996jy}. Large NG can also arise if reheating after inflation is inhomogeneous, varying from one causal (Hubble) region to another---as studied in this work.

In this paper we study the NG caused by effects of the Higgs boson of the Standard Model (SM) of particle physics during reheating~\cite{Ichikawa:2008ne,Choi:2012cp,DeSimone:2012gq,Cai:2013caa,Fujita:2016vfj,Lu:2019tjj}. Specifically, the Higgs boson can be responsible for inhomogeneous reheating and the corresponding generation of (non-Gaussian) density perturbations. Keeping our discussion as general as possible, we remain agnostic as to the inflationary model, as long as it reheats via perturbative inflaton decay to SM particles coupled to the Higgs boson. Our scenario is minimal since we do not introduce any new particles beyond the SM apart from the inflaton itself. Here the Higgs is not the inflaton; instead it is a light spectator field with vastly subdominant energy density compared to the inflaton. 

We also assume that quantum fluctuations of the inflaton yield the nearly scale invariant and highly Gaussian spectrum of density perturbations characteristic of the single field slow-roll inflation paradigm and consistent with observations of the CMB. In addition to perturbations associated with the quantum fluctuations of the inflaton, the effects of the Higgs boson on reheating can induce a independent (i.e.\ uncorrelated) contribution to the perturbation spectrum. We demonstrate that the density perturbations associated with reheating can induce a large NG signal, even when the amplitude of the perturbations is much smaller than that of perturbations associated with the quantum fluctuations of the inflaton.

The Higgs field acquires large quantum fluctuations during inflation. As a result, the Higgs field has different values in parts of the Universe which become causally disconnected from one another during inflation~\cite{Dvali:2003ar}. Since the Higgs imparts mass to SM particles, reheating can be delayed until the SM masses become lower than the inflaton mass, as we showed in Ref.~\cite{Freese:2017ace}. Further, spatial fluctuations in the Higgs values lead to spatial fluctuations in particle masses, and the reheating process becomes inhomogeneous. Inhomogeneous reheating caused by the probabilistic behavior of a light scalar field, in this case the Higgs boson, is called {\it modulated reheating}.

In Ref.~\cite{Litsa:2020rsm} we computed the amplitude of Higgs-induced temperature anisotropies and used CMB data to constrain model parameters. In this work, we show that the non-linearity of Higgs-modulated reheating processes can be the cause of significant NG in the resulting density perturbation spectrum. We find that CMB bounds on NG set by the \textit{Planck} measurements~\cite{Planck:2019kim} provide even more powerful constraints than those obtained in Ref.~\cite{Litsa:2020rsm} and, thus, exclude the possibility of perturbations from Higgs-modulated reheating providing the dominant contribution to the observed power spectrum of temperature anisotropies. By connecting primordial NG to SM parameters such as the Higgs self-coupling, we demonstrate that future NG signals can be used to probe the evolution of the Higgs field during inflation, thereby probing its potential over energies that are otherwise inaccessible.
 
Previous works have calculated the NG signal associated with modulated reheating due to the particle masses induced by a light spectator SM Higgs boson~\cite{Fujita:2016vfj,Lu:2019tjj}. While a detailed comparison is beyond the scope of this work, we note that previous calculations have utilized the mean field approach when considering the stochastic dynamics of spectator Higgs during inflation and the $\delta N$ formalism for the associated spectrum of density perturbations. Alternatively, we modify the approach of Ref.~\cite{Dvali:2003em} to track the growth of density perturbations from the end of inflation on superhorizon scales in causally disconnected Hubble patches. To set an initial condition for the post-inflationary evolution of the Higgs field in each Hubble patch, we draw from the equilibrium distribution of field values associated with the stochastic dynamics of light spectator fields during inflation. This {\it patch-by-patch} method typically results in NG signals which yield constraints on Higgs-induced temperature anisotropies that are significantly more stringent compared to previous calculations.
 
The rest of the paper is outlined as follows. In Sec~\ref{sec:calc_dens_pert}, we describe our calculation of the density perturbations associated with modulated reheating. We then describe the corresponding temperature anisotropies in the CMB in Sec.~\ref{sec:temp_and_ng}. In Sec.~\ref{sec:conclusions}, we conclude with a brief discussion of our results. 
 
\section{Calculation of density perturbations}
\label{sec:calc_dens_pert}
\subsection{Higgs field fluctuations}
\label{sec:Higgs_field_fluct}

We fix the background value of the Higgs doublet and its potential as\footnote{Assuming the central values of the top quark and Higgs masses, the SM Higgs potential becomes unstable at inflation scales $\gtrsim 10^{11}{\rm\,GeV}$~\cite{Degrassi:2012ry,Buttazzo:2013uya,Bezrukov:2012sa,Enqvist:2014bua}. Various mechanisms have been proposed to stabilize the electroweak (EW) vacuum at the inflation scale, including couplings between the Higgs and the inflaton (for example, see Ref.~\cite{Lebedev:2012sy}). However, the SM Higgs potential can also maintain stability up to $\sim 10^{15} {\rm GeV}$ for a top quark mass $3 \sigma$ below the central value. In order to emphasize the effects of the Higgs on the temperature fluctuations observed in the CMB without direct couplings to the inflaton, we assume the latter scenario and will consider the former in future work.}
\begin{eqnarray}
\Phi = \frac{1}{\sqrt{2}}\left(\bfrac{0}{h}\right)\, ,\,
V_{\rm H}(h) = \frac{\lambda}{4}\left(\Phi^\dag\Phi - \frac{\nu^2}{2}\right)^2 \approx \frac{\lambda}{4}\,h^4 \, , ~\label{eq:higgs_potential}
\end{eqnarray}
where $\nu = 246\,$GeV, $\lambda$ is the Higgs self-coupling, and $h$ is a real scalar field.

Due to quantum fluctuations of the Higgs field during inflation, super-horizon Higgs modes follow a random walk during the final stages of inflation. As a result, the Probability Density Function (PDF) describing the Higgs field at the end of inflation is~\cite{Starobinsky:1994bd}
\begin{equation}
\label{eq:PDFi}
 {g}_{\rm eq}(h) \!=\! \left(\frac{32\pi^2\lambda_I}{3}\right)^{1/4}\!\frac{1}{\Gamma (1/4)H_I}\!\exp\left(\!-\frac{2\pi^2 \lambda_I h^4}{3H_I^4} \!\right)\,,
\end{equation}
where $\Gamma(1/4)\approx 3.625$ and $\lambda_I$, $H_I$ are the self-coupling and the Hubble rate at the end of inflation, respectively.\footnote{We neglect the bare Higgs mass compared to the self-interaction term and consider a stabilized Higgs potential $\lambda_I>0$ during inflation.} Although the method we outline can generally be applicable to any inflationary model, we focus on models in which the Hubble scale at the start of reheating is roughly equal to the inflaton mass, i.e.\ $H_I \simeq m_{\phi},$ a choice typically made since it applies to many single field models in which observables (e.g.\ the tensor-to-scalar ratio) are well within reach of next-generation CMB experiments.

\subsection{Patch-by-patch method}
\label{sec:patch_by_patch}

In our novel approach, we treat each Hubble patch as a homogeneous {\it separate Universe}~\cite{Wands:2000dp}, where the energy densities in the $i$-th patch for the inflaton ${\rho}^i_{\phi}$ and for radiation ${\rho}^i_{r}$ evolve as
\begin{eqnarray}
\dot{\rho}^i_{\phi} &=& -3H^i{\rho}^i_{\phi}-{\Gamma}^i_{\phi} {\rho}^i_{\phi}\,,\label{eq:Boltzmann_m} \\
 \dot{\rho}^i_r &=& -4H^i{\rho}^i_r+{\Gamma}^i_{\phi} {\rho}^i_{\phi}\,,\label{eq:Boltzmann_r}\\
H^i  &=& \sqrt{\frac{8\pi G}{3}\left({\rho}^i_{\phi}+{\rho}^i_r\right)}\,. \label{eq:Hubble}
\end{eqnarray}
Here, $H^i$ is the Hubble scale in the $i$-th patch, $G$ is Newton's constant, and a dot denotes a derivative with respect to cosmic time. The decay rate of the inflaton (matter) into SM Higgsed fermions (radiation) is~\cite{Freese:2017ace}
\begin{equation}
\label{eq:decayrate}
\Gamma ^i _{\phi} = \Gamma_0\,\left(1-\frac{2y^2(h^i)^2}{m_{\phi}^2}\right)^{3/2}\Theta\left(m_\phi^2 - 2y^2(h^i)^2\right)\;,
\end{equation}
where $\Gamma_0$ is the (unblocked) decay rate of the inflaton for $h^i \to 0$ and the second factor corresponds to a Yukawa-like coupling\footnote{Yukawa couplings of a SM singlet inflaton to SM fermions can be provided for by interactions involving new degrees of freedom with dynamics that are only relevant at energies well above the inflation scale. For instance, a dimension-5 effective operator of the form $\phi S \bar f f / \Lambda$ could be generated with a coupling to a scalar field $S$, which carries the relevant SM quantum numbers to preserve gauge invariance. Rather than the SM Higgs boson, $S$ could be a particle with identical SM charges and a mass sufficiently heavy to suppress quantum fluctuations during inflation. For a characteristic scale of the new dynamics $\Lambda$ and a vacuum expectation value $\langle S \rangle$ above the inflation scale, the relevant Yukawa coupling would be $y_\phi \propto \langle S \rangle / \Lambda$.} between the inflaton and fermions. The fermion mass in each Hubble patch is determined by the local value of the Higgs field $h^i$ as $m_f^i = y\,|h^i|/\sqrt{2}$, where $y$ is the associated SM Yukawa coupling. After inflation, the Higgs field in each Hubble patch evolves as~\cite{Enqvist:2013kaa,Enqvist:2014bua}\footnote{We assume that the Higgs oscillates slowly relative to the Hubble rate and do not explore the case in the opposite limit where the Higgs field oscillates more rapidly. In Ref.~\cite{Litsa:2020rsm}, we show that the results for the temperature fluctuation amplitude are nearly identical by assuming either slow or rapid Higgs oscillations. We have verified that taking either limit for Higgs oscillations yields similar results for the NG and only present one case for clarity.}
\begin{equation}
\label{eq:HiggsEOM}
\ddot{h}^i+3H^i\dot{h}^i+\lambda_I \left(h^i\right)^3 = 0\,.
\end{equation}

Considering a representative sample of $N_p=200$ causally disconnected Hubble patches, we numerically evolve Eqs.~\eqref{eq:Boltzmann_m}-\eqref{eq:HiggsEOM} in each patch, randomly drawing the initial condition for the Higgs field from the PDF in Eq.~\eqref{eq:PDFi}. The distributions of the inflaton and radiation energy densities are calculated at each time and the averages of the density distributions over all patches are given by $\bar{\rho}_s \equiv \sum_{i=1}^{N_p}p_i\,\rho^i_s/\mathcal{N}$ ($s \in \{ \phi, r \}$), where the weight $p_i$ is extracted from $ {g}_{\rm eq}(h)$ and $\mathcal{N} = \sum_{i = 1}^{N_p} p^i$. Density perturbations are then defined as $\delta^i_s \equiv \rho^i_s/\bar{\rho}_s - 1$. Since we assume the Higgs PDF remains in equilibrium as all observable scales exit the horizon, the spectrum of density perturbations is scale invariant in the pure de-Sitter limit. 

\begin{figure}[t] 
\includegraphics[width=1\linewidth]{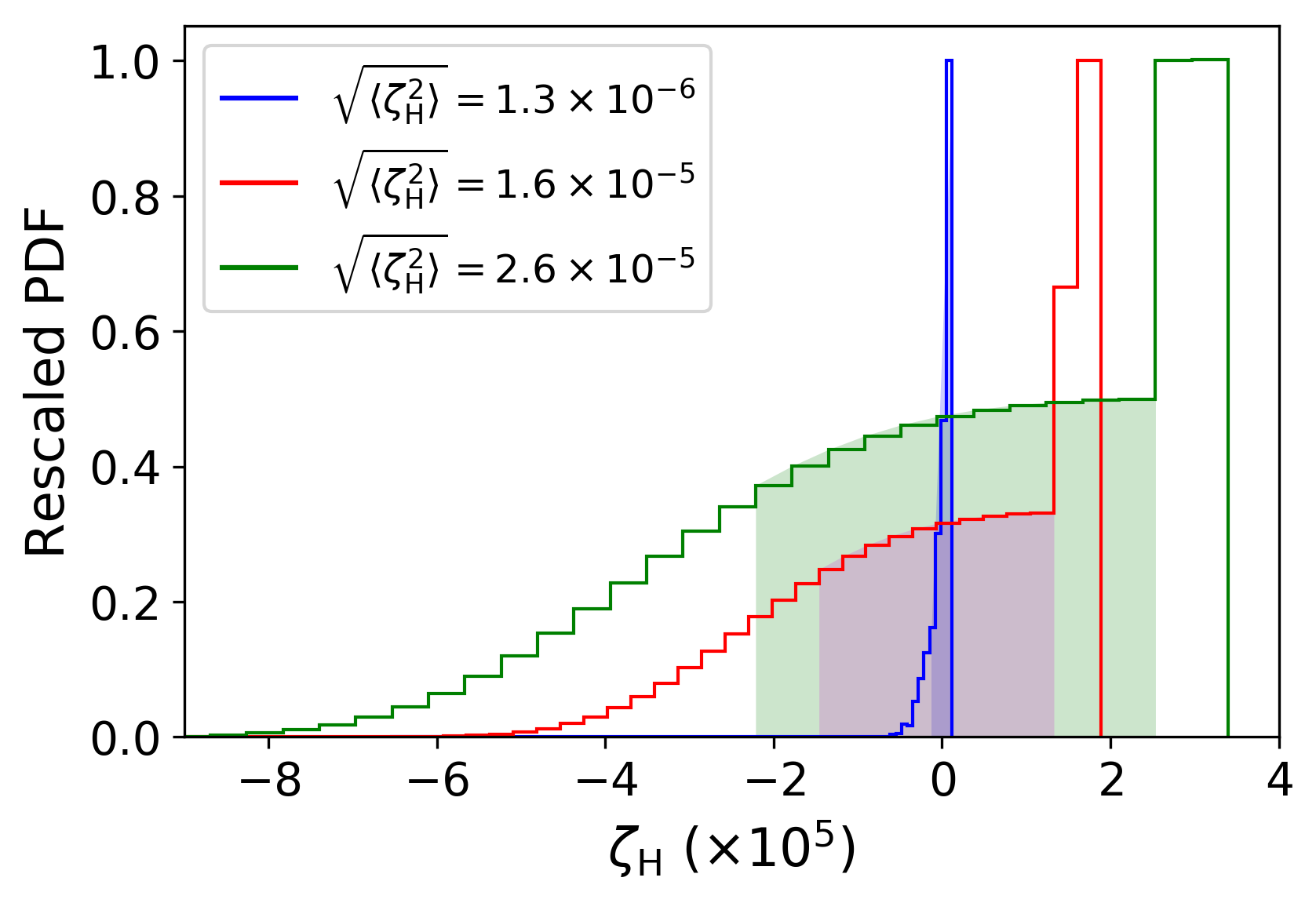} 
\caption{The PDF of the Bardeen parameter $\zeta_{\rm H}$ at $N=1,3,8$ $e$-folds after the end of inflation (blue, red and green lines, respectively), for $y = 10^{-2}$, $\Gamma_0 = 10^{-2}\,m_{\phi}$, and $\lambda_I = 10^{-2}$. Note that we only consider density perturbations from Higgs-modulated reheating, as defined in Eq.~\eqref{eq:zeta}. For better visualization, the PDFs are rescaled by their maximum value (and hence plotted with values between 0 and 1). The colored regions show the $1 \sigma$ regimes for the three distributions and the associated root mean squared values are shown in the legend. The step-like behavior is a numerical artifact arising from the binning process.} 
\label{fig1}
\end{figure}

The gauge-invariant Bardeen parameter~\cite{Bardeen:1980kt} produced by Higgs effects in each patch, $\zeta_{\rm H}^i$, is obtained by solving~\cite{Dvali:2003em, Litsa:2020rsm}
\begin{eqnarray}
\label{eq:delta_Hubble_N}
\dot{\Phi}^i_{\rm H} &=& -H^i\Phi_{\rm H}^i -\frac{4\pi G}{3H^i}\left(\bar{\rho}_{\phi} \delta^i_{\phi} + \bar{\rho}_r \delta^i_r\right)\,,\\
\label{eq:zeta}
\zeta_{\rm H}^i &=& \Phi^i_{\rm H}-\frac{\bar{\rho}_{\phi}\delta^i_{\phi}+\bar{\rho}_r\delta^i_r}{3\bar{\rho}_{\phi}+4\bar{\rho}_r}\;,
\end{eqnarray}
where $\Phi_{\rm H}^i$ is the gravitational potential perturbation. Fig.~\ref{fig1} shows the PDF of the Bardeen parameter $g(\zeta_{\rm H})$ at $N=1, 3, 8$ $e$-folds after the end of inflation for $y = 10^{-2}$, $\Gamma_0 = 10^{-2}\,m_{\phi}$, and $\lambda_I = 10^{-2}$. For visualisation, each PDF is divided by its maximum values $ {g}_{\rm max}(N=1) \simeq 6.7\times 10^{5}$, $ {g}_{\rm max}(N=3) \simeq 6.0\times 10^4$, and $ {g}_{\rm max}(N=8) \simeq 2.6\times 10^4$. The PDF initially broadens with increasing $N$, but no longer changes much once $N\gtrsim 8$ (with only percent level changes of the standard deviation $\sqrt{\langle \zeta_{\rm H}^2 \rangle}$ at later times).

The (local) NG of the perturbation spectrum corresponding to the {\it final} Bardeen parameter in each Hubble patch, $\zeta_f^i$, is quantified by the non-linearity parameter $f_{\rm NL}$, defined via
\begin{equation}
\label{eq:fnl_def}
{g}(\zeta_f) =  {g}_G(\zeta_f) + \frac{3}{5} f_{\rm NL}\left[ {g}^2_G(\zeta_f) - \braket{ {g}^2_G(\zeta_f)}\right]\;,
\end{equation}
where $g_{G}(\zeta_f)$ is a Gaussian PDF with mean $\langle{g_G^{\,}}(\zeta_f)\rangle$ and variance $\langle{g_G^2}(\zeta_f) \rangle $, and $\langle ...\rangle$ denotes averaging across all patches. Given the (near) scale invariance of both perturbations arising from Higgs-modulation effects and those associated with the quantum fluctuations of the inflaton, we can make an order of magnitude estimate for the non-linearity parameter~\cite{Bartolo:2004if}
\begin{equation}
\label{eq:fnl}
f_{\rm NL}\approx \frac{5}{18} \frac{\mathcal{S}}{\zeta_{\rm rms}^4} \, ,
\end{equation}
where $\zeta_{\rm rms}^2 \equiv \left \langle g^2 \left( \zeta_f \right) \right \rangle$ and the skewness of $ {g}(\zeta_f)$ is
\begin{equation}
\label{eq:S_def}
\mathcal{S} \equiv \left \langle { {g}^3\left(\zeta_f\right)}\right \rangle  = \frac{1}{\mathcal{N}} \sum\limits_{i = 1}^{N_p} p^i\left[\zeta_f^i - \bar{\zeta}_f\right]^3 \,  .
\end{equation}

\begin{figure}
		\includegraphics[width=1.0\linewidth]{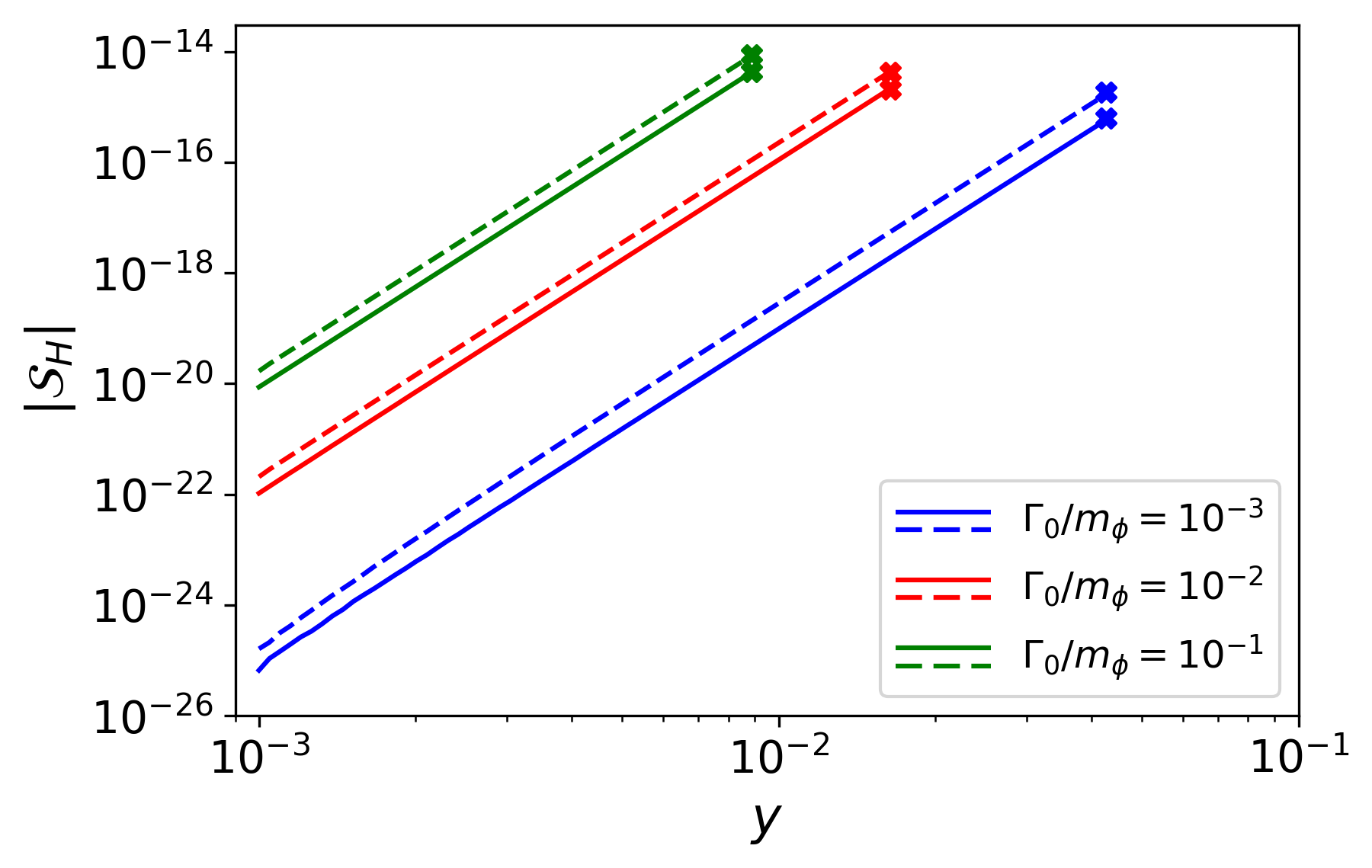} 
		\caption{The skewness ${\cal S}_H$ of the reheating-induced curvature-perturbation spectrum, as a function of the Yukawa coupling $y$, for different values of the unblocked inflaton decay rate $\Gamma_0$ (see Eq.~\eqref{eq:decayrate}). The latter is expressed in units of the inflaton mass, which we take to be equal to the Hubble scale at the end of inflation. Solid curves correspond to initial Higgs values drawn from the distribution of Eq.~\eqref{eq:PDFi}, while dashed curves correspond to a Gaussian initial PDF for the Higgs field.
		} 
	\label{fig:SH} 
\end{figure}

In principle, contributions to the non-linearity of the perturbation spectrum arising from both the quantum fluctuations of the inflaton and inhomogeneous reheating should be taken into account. However, in order to simplify our calculations and elucidate the role of Higgs-modulated reheating on the generation of primordial NG, we assume that the spectrum of perturbations from inflaton fluctuations is highly Gaussian, so that the NG signal is dominated by the reheating dynamics. With this assumption, the skewness ${\cal S}$ in Eq.~\eqref{eq:S_def} is solely determined by the skewness of the Higgs-induced density perturbation distribution $\mathcal{S}_{\rm H}$. In Fig.~\ref{fig:SH}, the solid curves show the skewness as a function of the Yukawa coupling $y$ and the unblocked inflaton decay rate $\Gamma_0$, with the initial Higgs values sampled from the PDF of Eq.~\eqref{eq:PDFi}.

We can estimate the contribution to the skewness of the PDF of density perturbations from Higgs-modulated reheating for the most relevant parameter space discussed in the next section by fitting our numerical results for $m_\phi = H_I$ and $\lambda_I=10^{-2}$,
\begin{equation}
\label{eq:SH}
\left | \mathcal{S}_{\rm H}  \right | \simeq 625.0 \,\left ( \frac{\Gamma_0}{m_{\phi} } \right)^{2.9} \, y^{5.7}\,,
\end{equation}
which is valid over $10^{-3} \leq y \leq 1$ and $10^{-7} \leq \Gamma_0/m_{\phi} \leq 10^{-1}$. The skewness is enhanced for inflaton couplings to SM fermions with larger Yukawa couplings and for larger perturbative decay widths. As we discuss in detail in Ref.~\cite{Litsa:2020rsm}, the effects of the Higgs on reheating are more significant for the respective larger fermion masses in each Hubble patch and the faster (unblocked) decay rate of the inflaton. 

\section{Constraints on Temperature Anisotropies}
\label{sec:temp_and_ng}

\begin{figure}[t]
\includegraphics[width=1.\linewidth]{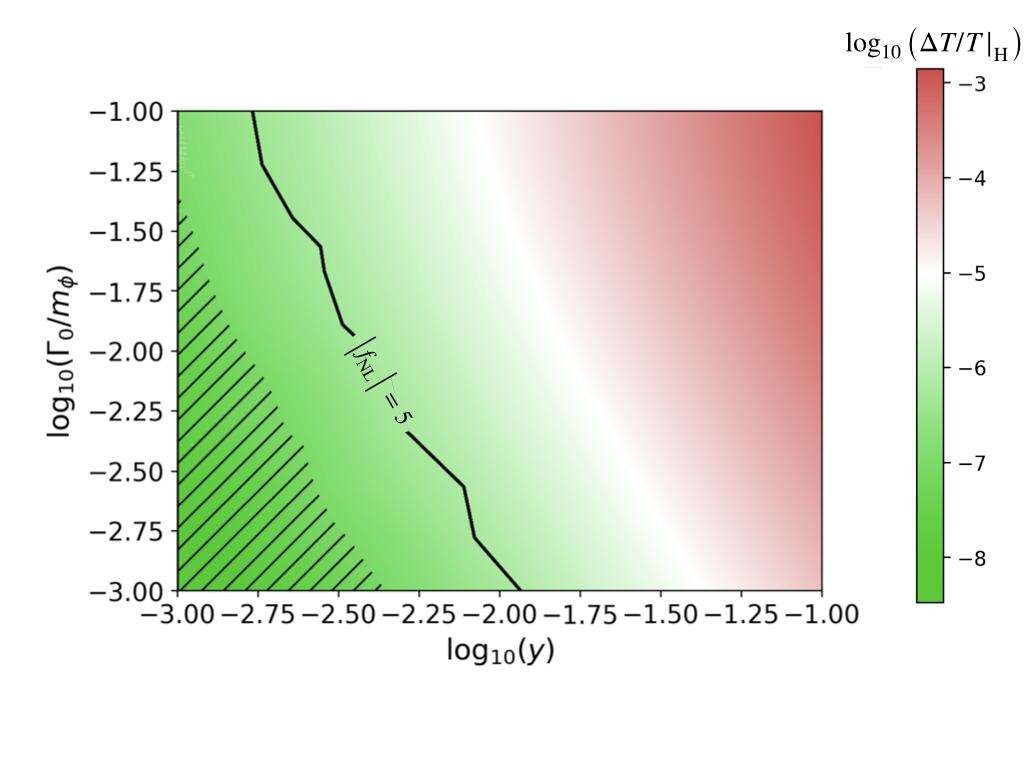} 
\setlength{\abovecaptionskip}{-20pt}
\caption{Parameter constraints from requiring that the amplitude~\cite{Litsa:2020rsm} and NG of temperature fluctuations from Higgs-modulated reheating do not exceed CMB observations.  $\Gamma_0$ is the unblocked inflaton decay rate; $y$ is the Yukawa coupling of SM particles to the Higgs; and ${\cal T}_{\rm H}\equiv \Delta T/T|_{\text{H}}$ is the overall temperature fluctuation induced by the Higgs modulation of reheating~\cite{Litsa:2020rsm}. We take the Higgs self-coupling and the Hubble parameter at the end of inflation to be $\lambda_I = 10^{-2}$ and $H_I = m_{\phi}$, respectively. The red region corresponds to Higgs-induced temperature inhomogeneities ${\cal T}_{\rm H}$ which are larger than those observed in the CMB. The white and green regions satisfy ${\cal T}_{\rm H} \lesssim 10 ^{-5}$ and are allowed by observations of the temperature fluctuation amplitude. The hatched region at ${\cal T}_{\rm H} \lesssim 10 ^{-7}$ indicates the regime in which Higgs effects cannot be observed, since the temperature fluctuations ${\cal T}_{\rm H}$ are smaller than the $\sim 1\%$ precision of \textit{Planck}. The black line labeled ``$\left | {f}_{\rm NL}\right | =5$'' represents the contour of constant ${f}_{\rm NL}$ corresponding to the lower limit set by the \textit{Planck} satellite; i.e.\ the region to the right of that black line has already been ruled out.} 
\label{fig2} 
\end{figure}

The PDF of temperature fluctuations $g({\cal T})$ on the largest angular scales observed in the CMB is derived from $g(\zeta_f)$ using the relation between temperature fluctuations and the final value of the Bardeen parameter in the $i$-th Hubble patch after the end of reheating~\cite{Liddle:1993fq, White:1997vi}
\begin{equation}
\label{eq:THi}
\mathcal{T}^i \equiv \frac{\Delta T}{T}\Big|^i = \frac{\zeta^i_f}{5}\,.
\end{equation}
Similar to the PDF of density perturbations, the variance of the associated PDF of temperature fluctuations is given by $\mathcal{T}_{\rm rms}^2 \equiv \left \langle g^2 \left( \mathcal{T} \right) \right \rangle$

In Fig.~\ref{fig2}, we show the results for our calculations of the temperature fluctuations arising from Higgs-modulated reheating while fixing $\lambda_I = 10^{-2}$ and $H_I = m_{\phi}$. The color contours show the amplitude of temperature fluctuations (see Ref.~\cite{Litsa:2020rsm} for details) when scanning over the parameters $(y,\Gamma_0)$. In the red region, the Higgs-induced perturbations are larger than what is observed in the CMB, $\mathcal{T}_{\rm H} \gtrsim \mathcal{T}_{\rm CMB}$, while in the green region $\mathcal{T}_{\rm H} \lesssim \mathcal{T}_{\rm CMB}$. Hence, both white and green regions are allowed by the amplitude of temperature fluctuations alone. The hatched region on the bottom left of Fig.~\ref{fig2} corresponds to ${\cal T}_{\rm H}\lesssim 10^{-7}$. Any contribution of Higgs effects to the total temperature fluctuation spectrum associated with this hatched region will not be detectable, being below the $\mathcal{O}(1\%)$ sensitivity of the \textit{Planck} satellite~\cite{Planck:2018jri}. We focus our attention on the green region and take the total temperature fluctuation amplitude at the largest angular scales to match the normalization of the observed power spectrum, ${\cal T}_{\rm rms} = \mathcal{T}_{\rm CMB} \sim 10^{-5}$. 

The region where $\mathcal{T}_{\rm H} \lesssim \mathcal{T}_{\rm CMB}$ can nonetheless lead to a large NG signal. The black line in Fig.~\ref{fig2} shows the parameter choices corresponding to $\left | {f}_{\rm NL}\right | = 5$, calculated using Eq.~\eqref{eq:fnl}. The current limit on local-type NG from the \textit{Planck} analysis reads $\left| f_{\rm NL} \right|  \lesssim 5$ and, thus, the region to the right of the line labeled ``$\left| f_{\rm NL} \right| = 5$'' is excluded. These results improve over the bounds we obtained in Ref.~\cite{Litsa:2020rsm} from solely using the amplitude of temperature fluctuations. For a given value of $y$, the bounds on $\Gamma_0$ from NG are nearly two orders of magnitude stronger than what is obtained from demanding $\mathcal{T}_{\rm H} \lesssim \mathcal{T}_{\rm CMB}$. Furthermore, the constraints from the NG signal imply that the dominant contribution to the observed power spectrum of temperature anisotropies cannot arise from Higgs-modulated reheating.

Here, we should note that our bound in Fig.~\ref{fig2} labeled ``$\left | {f}_{\rm NL}\right |  = 5$'' is very close to the limit below which Higgs effects on NG cannot be observed. In fact, regardless of its primordial value, small NG of $\left| f_{\rm NL} \right| \lesssim \mathcal{O}(1)$ will always be amplified to $\mathcal{O}(1)$ by secondary non-linear effects occurring before CMB decoupling~\cite{Bartolo:2004if}. The same effects make the primordial NG produced by quantum fluctuations of the inflaton in the standard slow-roll paradigm undetectable. As a result, a future detection of $\left| f_{\rm NL} \right| \sim \mathcal{O}(1)$ alone cannot confirm whether a NG signal has originated from inflation or reheating. At least for the scale invariant spectrum of density perturbations considered in this work, the parameter space of interest for further NG calculations and potential future observations of Higgs effects on NG is therefore limited.

Bounds on the combination of $\Gamma_0$ and $y$ from the NG of the temperature fluctuations produced by Higgs-modulated reheating can be used to constrain the reheat temperature for various SM decay channels of the inflaton. In general, lowering $\Gamma_0$ for a given choice of $y$ suppresses the amount of NG, as shown in Fig.~\ref{fig2}. However, $\Gamma_0$ depends on the reheat temperature as $T_{\rm reh} \propto \sqrt{\Gamma_0}$ and cannot be arbitrarily lowered without clashing with other early universe processes.

As an example, a fit for the dependence of $f_{\rm NL}$ on the parameters when the inflaton decays primarily to top quarks ($y=1$) is
\begin{equation}
|f_{\rm NL}| \simeq 5\left(\frac{10^{-2}}{\lambda_I} \right)^{0.9} \left( \frac{T_{\rm reh}}{5\times 10^{11}\, {\rm GeV}}\right)^{5.3} \left(\frac{10^{13} \, {\rm GeV}}{m_\phi}\right )^{2.7}\!.
\end{equation}
This fit is accurate at the $10\%$ level for $1\lesssim |f_{\rm NL}| \lesssim 100$.  For many inflation models, such as natural inflation with a cosine potential~\cite{Freese:1990rb}, $m_\phi$ typically lies in the range $10^{11-13} {\rm \, GeV}$. Assuming $\lambda_I=10^{-2}$, the requirement that $f_{\rm NL} \le 5$ can translate into an upper bound on the reheat temperature $T_{\rm reh} \le {\cal O}(10^{11}) \, {\rm GeV}$, which can conflict with the lower bound on $T_{\rm reh}$ arising from models of thermal leptogenesis (see e.g. Ref.~\cite{Giudice:2003jh,Buchmuller:2004nz}). More generally, the reheat temperature in similar inflationary models with $m_\phi \simeq H_I$ will become more constrained by the NG arising from Higgs-modulated reheating as the scale of inflation becomes smaller.

Our model links the reheating temperature (through $\Gamma_0$) and the Higgs potential (through $\lambda_I$) for a given particle species (through $y$). If future CMB experiments measure $f_{\rm NL}$ and hint at the value of $H_I$ through a detection of tensor modes, the method we presented would link the reheat temperature to the Higgs self-coupling for a given inflaton decay channel. When we incorporate limits on the reheat temperature from other early universe processes, like leptogenesis, a detection of $f_{\rm NL}$ would lead to a lower bound on $\lambda_I$, since lowering the value of $\lambda_I$ leads to both increased inhomogeneities $\Delta T/T$ and increased values of $f_{\rm NL}$. In principle, this would even allow us to probe the Higgs potential at inflationary energies and to constrain new physics in between the EW and inflation scales through the renormalization group (RG) running of $\lambda$.\footnote{We assume that the SM Yukawa couplings do no evolve significantly due to the RG running between the EW scale and inflation scale. On the other hand, the self-coupling of the SM Higgs $\lambda$ evolves significantly with scale and can receive relevant RG contributions from new physics above the EW scale.}

\subsection{Assumptions and Parameter Dependence}
\label{sec:assumtions} 
 
Given the constraints derived using the NG signal of Higgs-modulated reheating, it is natural to ask whether the specific choice of the Higgs PDF in Eq.~\eqref{eq:PDFi} is crucial. We repeated the computation by substituting Eq.~\eqref{eq:PDFi} by a Gaussian PDF with the same variance. The dashed lines in Fig.~\ref{fig:SH} show the corresponding results. In fact, a Gaussian initial PDF results in slightly higher skewness than the PDF given by Eq.~\eqref{eq:PDFi}, all other parameters being equal. Eq.~\eqref{eq:decayrate} shows that $\Gamma ^i _{\phi}$ is a non-linear function of $h^i$, which is the main contributor to the skewness   $\mathcal{S}_{\rm H}$. Hence, Eq.~\eqref{eq:SH} is robust, up to small corrections, for initial PDFs that differ in shape but have the same variance. 


The PDF in Eq.~\eqref{eq:PDFi} has been derived under the assumption of a pure de-Sitter space, which is only approximately true during inflation. A more realistic PDF for a light spectator field could depend on the exact inflationary evolution, even resulting in much larger field displacements than what found from Eq.~\eqref{eq:PDFi}~\cite{Hardwick:2017fjo}. Furthermore, the stochastic evolution of the Higgs field during inflation is closer to a four-dimensional random walk than a one-dimensional one, and as a result larger field values are expected~\cite{Adshead:2020ijf}. Taking the exact spectator evolution of the Higgs doublet into account would lead to tighter constraints for wide classes of inflationary models. Since our current paper attempts to provide a generic, conservative, and model-independent picture, we leave this analysis for future work.

\section{Conclusions}
\label{sec:conclusions}

In this paper, we show how to use CMB observations of primordial non-Gaussianity (NG) to probe the SM Higgs dynamics during inflation. We have uncovered a generic phenomenon that appears during reheating in any model of inflation where the inflaton decays directly to SM particles. It could also be relevant for reheating into a similarly Higgsed sector of new particles. During inflation, the Higgs boson obtains space dependent quantum fluctuations that lead to inhomogeneous reheating. Both the amplitude (studied in our previous paper~\cite{Litsa:2020rsm}) and NG of the associated temperature anisotropies are detectable in the CMB, with the strongest constraints arising from NG, as shown in Fig.~\ref{fig2}. Thus, perturbations from the effects of the Higgs during reheating cannot provide for the dominant contribution to the observed power spectrum of temperature anisotropies.
  
Our method allows for a number of generalizations and applications. Detailed information from the Higgs PDF beyond the de-Sitter approximation could further improve the constraints. Further progress can be also made by considering different shapes of the bispectrum generated through Higgs-modulated reheating, as well as higher-order correlations functions, which can be trivially computed in our formalism. Such improvements will allow us to use the full power of the CMB data acquired by \textit{Planck} and future experiments. 

Our results depend on the details of the Higgs dynamics during inflation and can be used to constrain unknown physics above the electroweak scale. This includes inferring the RG flow of the Higgs self-coupling at high energies and the presence of additional Planck-suppressed operators and stabilizing terms in the Higgs potential (see, for example~\cite{Fumagalli:2019ohr,Mantziris:2020rzh}). Extending our formalism to include scale information  will allow us to detect scale-dependent features, such as couplings of the Higgs to the inflaton, leading to a time-dependent effective Higgs mass during inflation. In anticipation of next generation CMB experiments which are able to better constrain primordial NG, the effects of Higgs-modulated reheating could provide a radically new window into particle physics at the inflation scale. 

\begin{acknowledgments}
We would like to thank Richard Easther, Sarah Shandera, and Spyros Sypsas for useful comments. AL, KF, and PS acknowledge support by the Vetenskapsr\r{a}det (Swedish Research Council) through contract No.~638-2013-8993 and the Oskar Klein Centre for Cosmoparticle Physics. KF~is grateful for support from the Jeff and Gail Kodosky Endowed Chair in Physics at the University of Texas, Austin. KF~acknowledges support from the Department of Energy through DoE grant DE-SC0007859 and the Leinweber Center for Theoretical Physics at the University of Michigan. EIS~acknowledges support from the Dutch Organisation for Scientific Research (NWO). EIS~acknowledges the support of a fellowship from ``la Caixa'' Foundation (ID 100010434) and from the European Union's Horizon 2020 research and innovation programme under the Marie Sk{\l}odowska-Curie grant agreement No.~847648. The fellowship code is LCF/BQ/PI20/11760021. EIS acknowledges support from IFAE, which is partially funded by the CERCA program of the Generalitat de Catalunya. LV~acknowledges support from the NWO Physics Vrij Programme ``The Hidden Universe of Weakly Interacting Particles'' with project No.~680.92.18.03 (NWO Vrije Programma), which is (partly) financed by the Dutch Research Council (NWO), as well as support from the European Union's Horizon 2020 research and innovation programme under the Marie Sk{\l}odowska-Curie grant agreement No.~754496 (H2020-MSCA-COFUND-2016 FELLINI). KF and LV would like to thank Perimeter Institute, where this line of research was started, for hospitality (KF is supported by the Distinguished Visitors Research Chair Program). The work of PS is partially supported by the research grant ``The Dark Universe: A Synergic Multi-messenger Approach'' No.~2017X7X85K under the program PRIN 2017 funded by the Ministero dell'Istruzione, Universit{\`a} e della Ricerca (MIUR), and by the ``Hidden'' European ITN project (H2020-MSCA-ITN-2019//860881-HIDDeN).
\end{acknowledgments}

\bibliography{HiggsPert}

\end{document}